\documentclass[jmp,graphicx]{revtex4-1}

\usepackage{ amssymb }
\usepackage{amsmath}
\usepackage{breqn}
\usepackage[utf8]{inputenc}  
\usepackage[T1]{fontenc}
\draft 

\begin{document}


\title{Solving the Riccati Equation} 



\author{Everardo Rivera-Oliva}
\email{everardo.rivera@cinvestav.mx}

\affiliation{{ 
Departamento de Física\\\
Centro de Investigación y de Estudios Avanzados del Instituto Politécnico Nacional\\\
P.O. box 14-740 C.P. 07000 Ciudad de México;México}
}


\date{\today}

\begin{abstract}
In this study, the Riccati equation is resolved using the generalized recursive integrating factor method. By applying a non-linear transformation to the dependent variable $y(x)$ of the Riccati equation, a second-order linear differential equation is derived for a variable $u(x)$ that is related to $y(x)$ through the aforementioned transformation. The second-order differential equation is then addressed using the aforementioned integrating factors method to derive the general solution for $u(x)$, which is subsequently transformed back to obtain the general solution for $y(x)$, thereby resolving the Riccati equation. The general solution to the Riccati equation is presented, followed by solving a few illustrative application examples.
\end{abstract}


\maketitle 


\section{Introduction}
The Riccati equation, a non-linear ordinary differential equation, was discovered by Jacopo Riccati in \cite{riccati1724animadversiones} in 1724, over three centuries ago. This equation is prevalent in numerous physical phenomena. For example, it is encountered in several Newtonian dynamic problems, especially central potential problems, as discussed in \cite{PhysRevE.65.047602}. In Quantum Mechanics, it serves as a method to analyze the one-dimensional Schrödinger equation as explored in \cite{fraga1999schrödinger,Schuch_2014}. Furthermore, in Cosmology, it manifests in the form that the Friedmann equations take when modeling a spatially homogeneous and isotropic universe, as investigated in \cite{valerio}, among other areas of physics where this equation models a physical phenomenon.\\
Although the Riccati equation was discovered three centuries ago, a comprehensive method to obtain a general solution remains elusive, despite various advances. As outlined in standard texts on differential equations (see \cite{arfken2013mathematical,zill2016differential,butkov1968mathematical}), the Riccati equation can be transformed into a second-order linear differential equation with variable coefficients. For specific instances of the Riccati equation, it is feasible to derive a general solution using this transformation method when the resulting second-order equation corresponds to one with known solutions. Additionally, a standard method of quadrature is detailed in fundamental differential equations literature (see \cite{arfken2013mathematical,zill2016differential,butkov1968mathematical}), which facilitates the determination of the general solution to the Riccati equation if a particular solution is available. Nonetheless, the process of identifying a particular solution that satisfies the Riccati equation is often more reliant on conjecture than on systematic procedure.\\
In this study it is found the general solution to the Riccati equation through the employment of the generalized recursive integrating factors method. By implementing a transformation on the Riccati equation variable $y(x)$, a second-order linear differential equation is formulated for a new variable $u(x)$, which is correlated to $y(x)$ via the previously mentioned transformation. The resolution of this equation for $u(x)$ can be achieved utilizing the generalized integrating factors technique, which addresses second-order linear equations, as demonstrated in \cite{riveraoliva2025solvinglineardifferentialequations} where a comprehensive solution to such equations was established. Upon resolving the second-order equation for $u(x)$, it becomes feasible to revert to the original variable $y(x)$ by means of the inverse transformation, thereby achieving the solution to the Riccati equation.\\
The structure of this work is organized as follows: Section \ref{2} introduces the recursive method for resolving differential equations, closely adhering to \cite{riveraoliva2025solvinglineardifferentialequations}. Section \ref{4} examines the generalized recursive integrating factors method for solving second-order linear differential equations, also closely following \cite{riveraoliva2025solvinglineardifferentialequations}. In Section \ref{5}, the general solution to the Riccati equation is derived, accompanied by a few illustrative examples. Finally, Section \label{6} presents the conclusions and perspectives.

\section{Review of Recursive method to solve ordinary differential equations}\label{2}
In this section the recursive method to solve ordinary differential equations will be introduced by following closely a previous work \cite{riveraoliva2025solvinglineardifferentialequations}.\\
As outlined in standard texts on differential equations (for an introductory examination, refer to \cite{zill2016differential}; for exhaustive reviews, see \cite{arfken2013mathematical}, \cite{butkov1968mathematical}, and \cite{hassani2013mathematical}), an ordinary differential equation pertaining to the unknown function $y(x)$ is an equation comprising $y(x)$ and the derivatives of $y(x)$ with respect to the independent variable $x$. Typically, an ordinary differential equation for $y(x)$ is written as:
\begin{equation}\label{r1}
    F\left(y(x),\frac{dy}{dx},\frac{d^2y}{dx^2},\hdots,\frac{d^ny}{dx^n},x,f(x)\right)=0
.\end{equation}
in which $f(x)$ is designated as the inhomogeneous function of the equation, remaining independent of $y(x)$. The determination of the order of the differential equation as indicated by Eq.\eqref{r1} is contingent upon the order of the highest derivative of $y(x)$ that is encompassed therein. The categorization of the various types of ordinary differential equations represented by Eq.\eqref{r1} is based on the properties satisfied by $F(y,y',...,y^{(n)},f)$.\\
A comprehensive resolution strategy for differential equations of the form delineated in Eq.\eqref{r1} is currently unavailable; solutions are accessible solely for specific instances of the functional form $F(y,y',...,y^{(n)},f)$ (cf. \cite{zill2016differential,arfken2013mathematical,butkov1968mathematical}). Within this context, a recursive methodology is proposed to address differential equations characterized by Eq.\eqref{r1}. This recursive approach necessitates solving for $y(x)$ in terms of $y(x)$, implying that the solution $y(x)$ is formulated as a function of itself and its derivatives. The method involves transforming $F(y,y',...,y^{(n)},f)$ into a known and tractable differential equation $H(y,y',...,y^{(m)},g)$ that is sufficiently straightforward to be resolved for $y(x)$:
\begin{equation}
    F(y,y',...,y^{(n)},f)=H(y,y',...,y^{(m)},g)
.\end{equation}
where $H(y,y',...,y^{(m)},g)$ denotes a differential equation for $y(x)$, which is recognized to be solvable. This is achieved at the expense that the inhomogeneous function within $H(y,y',...,y^{(m)},g)$ does not remain independent of $y(x)$ or its derivatives; rather, it is expressed as a function dependent on them:
\begin{equation}
    g=g(y,y',..,y^{(k)},x)
.\end{equation}
As a result, the solution for $y(x)$ obtained through this methodology will be expressed as a function of $y(x)$ itself together with its derivatives, exhibiting a recursive nature delineated by:
\begin{equation}\label{r2}
    y(x) = G\left(y(x),\frac{dy}{dx},\frac{d^2y}{dx^2},\hdots,\frac{d^ny}{dx^n},x,f(x)\right)
.\end{equation}
such that:
\begin{equation}\label{r3}
    F\left(G,\frac{dG}{dx},\frac{d^2G}{dx^2},\hdots,\frac{d^nG}{dx^n},x,f(x)\right)=0
.\end{equation}
Equation \eqref{r2} represents a recursive equation to $y(x)$ and serves as a solution to the ordinary differential equation specified in Eq.\eqref{r1}. By methodically applying recursive substitutions of the recursive relationship for $y(x)$ into $y(x)$, the explicit solution for Eq.\eqref{r1} can be determined:
\begin{equation}
    y(x)=G(G(G...(G(x))...))
.\end{equation}
In general, the recursive relationship may necessitate an infinite series of iterations, with no guarantee of convergence to a universal solution or even convergence of the function itself. The approach proves to be of practical significance when a discernible pattern within the recursion can be identified, thereby facilitating the derivation of an explicit expression for $y(x)$. An eminent attribute of this method is its applicability in conjunction with other well-established methodological frameworks for solving differential equations. In the current study, it is employed with additional techniques to resolve second-order ordinary differential equations.

\section{Review of Generalized recursive integrating factors}\label{4}
In this section the generalized recursive integrating factors method to solve second order linear differential equations will be introduced by following closely a previous work \cite{riveraoliva2025solvinglineardifferentialequations}.\\
The examination of second-order linear differential equations shall now commence. The canonical form of a second-order ordinary differential equation is presented as:
\begin{equation} \label{eq 2.1.1}
    \frac{d^2y}{dx^2}+p(x)\frac{dy}{dx}+q(x)y(x)=f(x)
.\end{equation}
The applicability of methods for solving Eq.\eqref{eq 2.1.1} is contingent upon the fulfillment of the conditions outlined in $p(x),q(x),f(x)$. Within the realm of physics, the Frobenius method, as delineated in \cite{+1873+214+235} regarding series solutions, represents one of the most prevalently utilized techniques. However, for certain differential equations, particularly those characterized by constant coefficients and Cauchy-Euler equations, distinct methods, as elucidated in \cite{zill2016differential,arfken2013mathematical,butkov1968mathematical}, are employed.
A principal advantage of the recursive method is its ability to convert ordinary differential equations into recognized solvable forms, thereby facilitating integration into established expressions. Moreover, this method can be effectively amalgamated with other solution techniques when advantageous. Consequently, the fundamental principle of the integrating factor method, relevant to first-order equations, is employed to reformulate Eq.\eqref{eq 2.1.1} as an exact differential equation by multiplying Eq.\eqref{eq 2.1.1} by two functions $\alpha(x),\beta(x)$:
\begin{equation} \label{eq 2.1.2}
    \alpha(x)\beta(x)\left(\frac{d^2y}{dx^2}+p(x)\frac{dy}{dx}+q(x)y(x)\right)=f(x) \alpha(x)\beta(x)
.\end{equation}
Observe that:
\begin{equation} \label{eq 2.1.3}
    \frac{d}{dx}\left(\alpha(x)\beta(x)\frac{dy}{dx}\right)=\alpha(x)\beta(x)\frac{d^2y}{dx^2}+\frac{d\alpha}{dx}\beta(x)\frac{dy}{dx}+\alpha(x)\frac{d\beta}{dx}\frac{dy}{dx}
.\end{equation}

\begin{equation}\label{eq 2.1.4}
    \frac{d}{dx}\left(\frac{d\alpha}{dx}\beta(x)y(x) \right) = \frac{d^2\alpha}{dx^2}\beta(x)y(x)+\frac{d\alpha}{dx}\frac{d\beta}{dx}y(x)+\frac{d\alpha}{dx}\beta(x)\frac{dy}{dx}
.\end{equation}
\begin{equation}\label{eq 2.1.5}
    \frac{d}{dx}\left(\alpha(x)\frac{d\beta}{dx}y(x)\right)=\frac{d\alpha}{dx}\frac{d\beta}{dx}y(x)+\alpha \frac{d^2 \beta}{dx^2}y(x)+\alpha(x)\frac{d\beta}{dx}\frac{dy}{dx}
.\end{equation}
By summing Eq. \eqref{eq 2.1.3} and Eq. \eqref{eq 2.1.5}, followed by the subtraction of Eq. \eqref{eq 2.1.4}, the resulting expression is obtained:
\begin{equation}\label{eq 2.1.6}
\begin{split}
    \frac{d}{dx}\left(\alpha(x)\beta(x)\frac{dy}{dx}\right)- \frac{d}{dx}\left(\frac{d\alpha}{dx}\beta(x)y \right)+\frac{d}{dx}\left(\alpha(x)\frac{d\beta}{dx}y\right) = \alpha(x) \beta(x)\frac{d^2y}{dx^2}+2\alpha(x)\frac{d\beta}{dx}\frac{dy}{dx}+\alpha(x)\frac{d^2\beta}{dx^2}y-\beta(x)\frac{d^2\alpha}{dx^2}y.
\end{split}
\end{equation}
To transform Eq.\eqref{eq 2.1.2} into an exact differential, it is imperative that the following conditions are fulfilled:
\begin{equation}\label{eq 2.1.7}
    \frac{d\beta}{dx}=\beta(x)\frac{p(x)}{2}
.\end{equation}
\begin{equation}\label{eq 2.1.8}
    \alpha(x)\frac{d^2\beta}{dx^2}-\beta(x)\frac{d^2\alpha}{dx^2}=\alpha(x)\beta(x)q(x)
.\end{equation}
In the event that $\alpha(x),\beta(x)$ is identified in accordance with the prescribed conditions, it becomes possible to construct a first integral:
\begin{equation} \label{eq 2.1.9}
    \alpha(x)\beta(x)\frac{dy}{dx}+\left(-\frac{d\alpha}{dx}\beta(x)+\alpha(x)\frac{d\beta}{dx}\right)y(x)= \int_0^x \alpha(t)\beta(t)f(t)dt + C_1
.\end{equation}
Following this development, a first-order linear differential equation is deduced. Upon reformulating this first-order ordinary differential equation into its canonical form, it is obtained:
\begin{equation}\label{eq 2.1.10}
    \frac{dy}{dx}+\frac{d}{dx}\ln\left( \frac{\beta}{\alpha}\right) y(x)=\frac{C_1}{\alpha(x)\beta(x)}+\frac{1}{\alpha(x) \beta(x)}\int_0^x\alpha(t)\beta(t)f(t)dt
.\end{equation}
The first-order ordinary differential equation delineated in Eq.\eqref{eq 2.1.10} can be resolved through the application of the integrating factor method or, alternatively, via the recursive method previously seen:
\begin{equation}\label{eq 2.1.11}
    \mu(x) = \frac{\beta(x)}{\alpha(x)}
.\end{equation}
then:
\begin{equation}\label{eq 2.1.12}
    \frac{d}{dx} \left(y(x)\frac{\beta}{\alpha(x)} \right)= \frac{C_1}{\alpha^2(x)}+\frac{1}{\alpha^2(x)}\int_0^x \alpha(t)\beta(t)f(t)dt
.\end{equation}
Consequently, integration of this equation yields $y(x)$:\begin{equation}\label{eq 2.1.13}
    y(x)=C_2\frac{\alpha(x)}{\beta(x)}+C_1\frac{\alpha(x)}{\beta(x)}\int_0^x \frac{dt}{\alpha^2(t)}+\frac{\alpha(x)}{\beta(x)}\int_0^x \frac{1}{\alpha^2(t)}\int_0^t \alpha(t')\beta(t')f(t')dt' dt
.\end{equation}
This represents the general solution to the second-order linear differential equation as specified in Eq.\eqref{eq 2.1.1}. Thus, given that $\alpha(x),\beta(x)$ are identifiable, one can obtain the general solution for $y(x)$ by employing this methodological approach, herein referred to as generalized recursive integrating factors. Upon resolving Eq.\eqref{eq 2.1.7}, $\beta(x)$ is obtained:
\begin{equation}\label{eq 2.1.14}
    \beta(x)=\exp\left({\int_0^x \frac{p(x)}{2}}\right)
.\end{equation}
By substituting Equation \eqref{eq 2.1.7} into Equation \eqref{eq 2.1.8}, an expression for $\alpha(x)$ is derived:
\begin{equation} \label{eq 2.1.15}
    \alpha(x)\left(\frac{1}{2}\frac{dp}{dx}+\frac{p^2}{4} \right)-\frac{d^2\alpha}{dx^2}=\alpha(x)q(x)
.\end{equation}
then in canonical form:
\begin{equation}\label{eq 2.1.16}
    \frac{d^2\alpha}{dx^2}+\alpha(x) \left[q(x)-\frac{1}{2}\frac{dp}{dx}-\frac{p^2}{4} \right]=0
.\end{equation}
Define $h(x)$ by:
\begin{equation}\label{eq 2.1.17}
    h(x)=-q(x)+\frac{1}{2}\frac{dp}{dx}+\frac{p^2}{4}
.\end{equation}
Consequently Eq.\eqref{eq 2.1.16} can be expressed as:
\begin{equation}\label{eq 2.1.18}
\frac{d^2 \alpha}{dx^2}-h(x)\alpha(x)=0
.\end{equation}
To obtain $\alpha(x)$, it is needed to resolve a second-order linear ordinary differential equation, whose general solution is not usually known. Hence, the recursion method shall be employed, focusing directly on the equation's second derivative while considering all other terms as inhomogeneous components. Subsequently, a first integral for $\alpha(x)$ is derived:
\begin{equation}\label{eq 2.1.19}
    \alpha(x) =C_2+C_1x+\int_0^x \int_0^{t_1} h(t_2)\alpha(t_2)dt_2dt_1
.\end{equation}
The objective is not to determine the general solution $\alpha(x)$, but rather to identify a particular solution. For the sake of simplicity, let us assume $C_1=0,C_2=1$. Consequently:
\begin{equation}\label{eq 2.1.20}
    \alpha(x)=1+\int_0^x \int_0^{t_1}h(t_2)\alpha(t_2)dt_2dt_1
.\end{equation}
By working out the first recursion it is obtained:
\begin{equation}\label{eq 2.1.21}
    \alpha(x)=1+\int_0^x\int_0^{t_1}h(t_2)dt_2dt_1+\int_0^x \int_0^{t_1}\int_0^{t_2}\int_0^{t_3}h(t_2)\alpha(t_4)dt_4dt_3dt_2dt_1
.\end{equation}
At this stage, a discernible pattern becomes apparent within the recursive procedure:
\begin{equation}\label{eq 2.1.22}
    \alpha(x)=1+\int_0^x\int_0^{t_1}h(t_2)dt_2dt_1+\int_0^x \int_0^{t_1}\int_0^{t_2}\int_0^{t_3}h(t_2)h(t_4)dt_4dt_3dt_2dt_1+\hdots
\end{equation}
In a compact notation $\alpha(x)$ can be expressed as follows:
\begin{equation}\label{eq 2.1.23}
\alpha(x)=1+\sum_{j\in 2 \mathbb{Z}^{+}}^{\infty}\prod_{i \in 2\mathbb{Z}^{+}}^{j} \int_0^{t_i}\int_0^{t_{i+1}} h(t_{i+2})dt_{i+2}dt_{i+1}
.\end{equation}
where $t_0=x$. It can be shown that Eq.\eqref{eq 2.1.22} fulfills the requirements of Eq.\eqref{eq 2.1.18} by calculating the second derivative of $\alpha(x)$:
\begin{equation}\label{eq 2.1.24}
    \frac{d\alpha}{dx} = \int_0^x h(t_2)dt_2+\int_0^{x}\int_0^{t_2}\int_0^{t_3}h(t_2)h(t_4)dt_4dt_3dt_2+\hdots
.\end{equation}
\begin{equation}\label{eq 2.1.25}
    \frac{d^2\alpha}{dx^2}=h(x)+h(x)\int_0^{x}\int_0^{t_3}h(t_4)dt_4dt_3+\hdots
.\end{equation}
\begin{equation}\label{eq 2.1.26}
    \frac{d^2\alpha}{dx^2}=h(x) \left[1+\int_0^x\int_0^{t_1}h(t_2)dt_2dt_1+\int_0^x \int_0^{t_1}\int_0^{t_2}\int_0^{t_3}h(t_2)h(t_4)dt_4dt_3dt_2dt_1+\hdots \right]
.\end{equation}
\begin{equation}\label{eq 2.1.27}
    \frac{d^2\alpha}{dx^2} = h(x)\alpha(x)
.\end{equation}
Thus, $\alpha(x)$ as delineated in Eq.\eqref{eq 2.1.21} fulfills the role of a solution to Eq.\eqref{eq 2.1.27}, thereby functioning as an integrating factor for Eq.\eqref{eq 2.1.1}. By rearranging the sequence of integration for $\alpha(x)$, a simplification of the expression is attainable:
\begin{equation}\label{eq 2.1.28}
    \alpha(x)=1+\int_0^x(x-t)h(t)dt+\int_0^x\int_0^t(x-t)(t-t_1)h(t)h(t_1)dt_1dt+\hdots
\end{equation}
By defining $x=t_0$ and definining a function $H(x)$ as follows:
\begin{equation}\label{eq 2.1.29}
    H(t_n)=(t_{n-1}-t_n)h(t_n)
.\end{equation}
then $\alpha(x)$ can be expressed as:
\begin{equation}\label{eq 2.1.30}
    \alpha(x) =1+\int_0^{t_0} H(t_1)dt_1+\int_0^{t_0} \int_0^{t_1} H(t_2)H(t_1)dt_2dt_1+\hdots
\end{equation}
However, Equation \eqref{eq 2.1.30} represents the expansion of the product-ordered exponential (refer to \cite{grossman1972non} for a comprehensive review). Consequently:
\begin{equation}\label{extra}
    \alpha(x) = \tau\left[\exp\left({\int_0^{x}H(t)dt}\right)\right]
.\end{equation}
Therefore $y(x)$ can be explicitly expressed as:

\begin{equation}\label{extra1}
\begin{split}
    y(x)=&C_2e^{-\int_{0}^{x} \frac{p(t)}{2}dt} \tau\left[e^{\int_0^{x}(x-t)\left(-q(t)+\frac{1}{2}\frac{dp}{dt}+\frac{p^2}{4} \right)dt}\right]\\ &+C_1e^{-\int_0^x \frac{p(t)}{2}dt} \tau\left[e^{\int_0^{x} (x-t)\left(-q(t)+\frac{1}{2}\frac{dp}{dt}+\frac{p^2}{4} \right)dt}\right] \int_0^x \frac{dt}{\tau\left[e^{-\int_0^{t}(t-t')\left(-q(t')+\frac{1}{2}\frac{dp}{dt'}+\frac{p^2}{4} \right)dt'}\right]^{2}}\\
    &+e^{-\int_0^x \frac{p(t)}{2}dt} \tau\left[e^{\int_0^{x} (x-t)\left(-q(t)+\frac{1}{2}\frac{dp}{dt}+\frac{p^2}{4} \right)dt}\right] \int_0^x\frac{\int_0^t f(t')e^{\int_0^{t'} \frac{p(z)}{2}dz} \tau\left[e^{\int_0^{t'}(t'-z)\left(-q(z)+\frac{1}{2}\frac{dp}{dz}+\frac{p^2}{4} \right)dz}\right]}{\tau\left[e^{-\int_0^{t}(t-t')\left(-q(t')+\frac{1}{2}\frac{dp}{dt'}+\frac{p^2}{4} \right)dt'}\right]^{2}}dt'dt.
\end{split}
\end{equation}
Equation \eqref{extra1} describes the general solution to the second-order linear differential equation as specified by Eq. \eqref{eq 2.1.1}. Though the notation may initially seem complex relative to the concise form of Eq. \eqref{eq 2.1.12} concerning integrating factors, it has the advantage of being expressed explicitly in terms of the functions $p(x),q(x)$ within the differential equation. Therefore, by employing the generalized integrating factor method alongside the recursive method, second-order linear differential equations can be systematically addressed in the general case.
\section{The Riccati Equation}\label{5}
The Riccati equation, a first-order non-linear differential equation, was first identified by Riccati in \cite{riccati1724animadversiones} in the year 1724, marking over three centuries since its initial discovery. In contemporary notation, the differential equation as presented in \cite{zill2016differential,arfken2013mathematical} is as follows:
\begin{equation}\label{Riccati 1}
    \frac{dy}{dx}=q_0(x)+q_1(x)y(x)+q_2(x)y^2(x).
\end{equation}
As articulated in foundational texts on differential equations, including \cite{arfken2013mathematical,butkov1968mathematical,zill2016differential}, Equation \eqref{Riccati 1} may be transformed into a second-order linear differential equation characterized by variable coefficients through the application of the subsequent transformation:
\begin{equation}\label{Riccati 2}
    y(x)=-\frac{u(x)'}{q_2(x)u(x)}.
\end{equation}
where $u(x)$ satisifies the following second order differential equation:
\begin{equation}\label{Riccati 3}
    \frac{d^2u}{dx^2}-\left(q_1(x)+\frac{d}{dx}\ln(q_2(x))\right)\frac{du}{dx}+q_2(x)q_0(x)u(x)=0.
\end{equation}
Thus, through the explicit resolution of Eq.\eqref{Riccati 3} for $u(x)$ and the employment of the transformation articulated in Eq.\eqref{Riccati 2}, one can obtain a solution to the Riccati equation. The subsequent terms are identified from Eq.\eqref{Riccati 3}:
\begin{equation}\label{Riccati 4}
    p(x)=-q_1(x)-\frac{d}{dx}\ln(q_2(x)).
\end{equation}
\begin{equation}\label{Riccati extra}
    q(x)=q_2(x)q_0(x).
\end{equation}
Thus, through the application of the generalized recursive integrating factors method \cite{riveraoliva2025solvinglineardifferentialequations}, one can determine the general solution for Eq.\eqref{Riccati 3}.Prior to elucidating the specific values for $\alpha(x),\beta(x)$, it is prudent to initially articulate all components in terms of these integrating functions. Given that the equation for $u(x)$ is homogeneous, the general solution is the following as specified by Eq.\eqref{eq 2.1.13}:
\begin{equation}\label{R0}
    u(x)=C_2 \frac{\alpha(x)}{\beta(x)}+C_1\frac{\alpha(x)}{\beta(x)}\int_0^x \frac{dt}{\alpha^2(t)}.
\end{equation}
In order to reconstruct $y(x)$, it is necessary to compute the logarithm of $u(x)$ followed by its differentiation. The $log(u(x))$ is provided as follows:
\begin{equation}\label{R1}
    \ln(u)=\ln \alpha(x)-\ln \beta(x)+\ln \left(C_2+C_1 \int_0^x \frac{dt}{\alpha^2(t)} \right).
\end{equation}
Subsequently its derivative is given by:
\begin{equation}\label{R2}
    \ln(u)'=\frac{\alpha'(x)}{\alpha(x)}-\frac{\beta'(x)}{\beta(x)}+\frac{C_1}{\alpha^2(x)} \frac{1}{C_2+C_1\int_0^x \frac{dt}{\alpha^2(t)}}.
\end{equation}
Renaming the integration constants appearing in Eq.\eqref{R2} the expression reduces into:
\begin{equation}\label{R3}
    \ln(u)'=\frac{\alpha'(x)}{\alpha(x)}-\frac{\beta'(x)}{\beta(x)}+\frac{1}{\alpha^2(x)} \frac{1}{C+\int_0^x \frac{dt}{\alpha^2(t)}}.
\end{equation}
By substituting Eq.\eqref{R3} into Eq.\eqref{Riccati 2} the general solution $y(x)$ for the Riccati equation is given by:
\begin{equation}\label{R4}
y(x) =\frac{\beta'(x)}{q_2(x)\beta(x)}-\frac{\alpha'(x)}{q_2(x)\alpha(x)}-\frac{1}{q_2(x)\alpha^2(x)}\frac{1}{C+\int_0^x \frac{dt}{\alpha^2(t)}}.
\end{equation}
Equation \eqref{R4} reveals that the general solution for the Riccati equation is comprised of a particular solution $y_p(x)$ that is independent of the integration constant and satisfies the Riccati equation itself (see \cite{zill2016differential,arfken2013mathematical}), as well as a commonly referred solution by quadrature $y_q(x)$ dependent on the constant of integration, satisfying a Bernoulli equation through a change of variable (see \cite{zill2016differential,arfken2013mathematical}). Consequently:
\begin{equation}\label{R5}
    y(x)=y_p(x)+y_q(x).
\end{equation}
where:
\begin{equation}\label{R6}
\begin{split}
    &y_p(x)=\frac{\beta'(x)}{q_2(x)\beta(x)}-\frac{\alpha'(x)}{q_2(x)\alpha(x)},\\
    &y_q(x)=-\frac{1}{q_2(x)\alpha^2(x)}\frac{1}{C+\int_0^x \frac{dt}{\alpha^2(t)}}.
\end{split}
\end{equation}
As evidenced in Eq. \eqref{R6}, the quadrature solution of the Riccati equation is solely dependent on a single integration constant, which is consistent with the nature of first-order differential equations. Despite the resolution of the equation through a second-order differential equation pertaining to $u(x)$, which incorporated two integration constants $C_1,C_2$, as illustrated in Eq. \eqref{R0}, the logarithmic transformation linking $u(x)$ with $y(x)$ in Eq. \eqref{R2} necessitates that these integration constants appear as a quotient for $y(x)$, rendering only $C_1/C_2$ significant. Consequently, it can be reformulated into a new constant $C$. Having derived the general solution to the Riccati equation in terms of the integrating factors $\alpha(x),\beta(x)$, it is imperative to determine explicit expressions for these functions in terms of the component functions presented in Eq. \eqref{Riccati 1}. According to Eq. \eqref{eq 2.1.14} and Eq. \eqref{Riccati 4}, the function $\beta(x)$ is given as follows:
\begin{equation}\label{Riccati 5}
\beta(x)=\frac{\exp\left({-\frac{1}{2}\int_0^xq_1(t)dt}\right)}{\sqrt{q_2(x)}}.
\end{equation}
To compute $\alpha(x)$, it is imperative to initially determine $h(x)$, as delineated in Eq.\eqref{eq 2.1.17}. Once $q(x),p(x)$ has been identified, the function $h(x)$ can subsequently be articulated as follows:
\begin{equation}\label{R10}
    h(x)=-q_2(x)q_0(x)-\frac{1}{2}\frac{dq_1(x)}{dx}-\frac{1}{2}\frac{d^2}{dx^2}\ln(q_2(x))+\frac{1}{4}q_1^2(x)+\frac{1}{2}q_1(x)\frac{d}{dx}\ln(q_2(x))+\frac{1}{4}\left(\frac{d}{dx}\ln(q_2(x))\right)^2.
\end{equation}
Subsequently we compute $H(x)$ function as follows:
\begin{equation}\label{R11}
    H(x)=(x-x')\left[-q_2q_0-\frac{1}{2}\frac{dq_1}{dx'}-\frac{1}{2}\frac{d^2 \ln q_2}{dx^2}+\frac{1}{4}q_1^2+\frac{1}{2}q_1\frac{d \ln q_2}{dx'}+\frac{1}{4}\left(\frac{d \ln q_2}{dx'}\right)^2\right].
\end{equation}
Consequently $\alpha(x)$ is given by the exponential ordered product of the integral of $H(x)$ as follows:
\begin{equation}\label{R12}
\begin{split}
    \alpha(x)=\tau\left[\exp\int_{0}^{x}\left((t'-t)\left(-q_2q_0-\frac{1}{2}\frac{dq_1}{dt}-\frac{1}{2}\frac{d^2 \ln q_2}{dt^2}+\frac{1}{4}q_1^2+\frac{1}{2}q_1\frac{d\ln q_2}{dt}+\frac{1}{4}\left(\frac{d \ln q_2}{dt}\right)^2\right)\right)dt\right].
    \end{split}
\end{equation}
Upon calculating $\alpha(x),\beta(x)$, expressions for $y_p,y_q$ can be derived in terms of the component functions of the Riccati equation. By inserting the values of $\alpha(x),\beta(x)$ into Eq.\eqref{R6}, the following result is obtained:
\begin{equation}\label{R13}
\begin{split}
    y_p(x)=&-\frac{1}{q_2(x)}\frac{d}{dx} \ln \left( \tau\left[\exp\int_{0}^{x}\left((t'-t)\left(-q_2q_0-\frac{1}{2}\frac{dq_1}{dt}-\frac{1}{2}\frac{d^2 \ln q_2}{dt^2}+\frac{1}{4}q_1^2+\frac{1}{2}q_1\frac{d\ln q_2}{dt}+\frac{1}{4}\left(\frac{d \ln q_2}{dt}\right)^2\right)\right)dt\right] \right)\\
    &-\frac{1}{2q_2(x)}\left[q_1(x)+\frac{d\ln q_2(x)}{dx} \right].
    \end{split}
\end{equation}
\begin{equation}\label{R14}
    \begin{split}
    y_q(x)=-\frac{\left(C+\int_0^{x} \left[\tau\left[\exp\int_{0}^{t}\left((z-y)\left(-q_2q_0-\frac{1}{2}\frac{dq_1}{dy}-\frac{1}{2}\frac{d^2 \ln q_2}{dy^2}+\frac{1}{4}q_1^2+\frac{1}{2}q_1\frac{d\ln q_2}{dy}+\frac{1}{4}\left(\frac{d \ln q_2}{dy}\right)^2\right)\right)dy\right]\right]^{-2}dt \right)^{-1}}{q_2(x) \left[\tau\left[\exp\int_{0}^{x}\left((t'-t)\left(-q_2q_0-\frac{1}{2}\frac{dq_1}{dt}-\frac{1}{2}\frac{d^2 \ln q_2}{dt^2}+\frac{1}{4}q_1^2+\frac{1}{2}q_1\frac{d\ln q_2}{dt}+\frac{1}{4}\left(\frac{d \ln q_2}{dt}\right)^2\right)\right)dt\right] \right]^2}.
    \end{split}
\end{equation}
It is evident that Eq.\eqref{R13} and Eq.\eqref{R14} are intricate and lengthy, but constitute the general solution to the Riccati equation. Practically, a more straightforward approach is to compute directly for $\alpha(x),\beta(x)$ and subsequently employ the transformation delineated in Eq.\eqref{Riccati 2} to derive the general solution.\\

To demonstrate the effectiveness of the derived general solution to the Riccati equation, a couple of illustrative examples shall be solved.\\
Consider the following Riccati equation:
\begin{equation}\label{a1}
    \frac{dy}{dx}=-\lambda\exp \left(\frac{x^2}{2}\right)+xy+\exp\left(-\frac{x^2}{2} \right)y^2.
\end{equation}
From here, the following functions are identified:
\begin{equation}\label{a2}
    q_0(x)=-\lambda\exp \left(\frac{x^2}{2}\right),\quad q_1(x)=x,\quad q_2(x)=\exp \left(-\frac{x^2}{2}\right).
\end{equation}
Subsequently, the functions $p(x),q(x)$ are constructed by following Eq.\eqref{Riccati 4} and Eq.\eqref{Riccati extra} as follows:
\begin{equation}\label{a3}
    p(x)=0,\quad q(x)=-\lambda.
\end{equation}
From Eq.\eqref{a3} and by following Eq.\eqref{eq 2.1.14} $\beta(x)$ is computed:
\begin{equation}\label{a4}
    \beta(x)=0.
\end{equation}
In the same way, following Eq.\eqref{eq 2.1.18} an equation for $\alpha(x)$ is computed:
\begin{equation}\label{a5}
    \frac{d^2 \alpha}{dx^2}=\lambda \alpha.
\end{equation}
The solution for Eq.\eqref{a5} is given by Eq.\eqref{extra} as follows:
\begin{equation} \label{a6}
    \alpha(x)=\exp \left(\sqrt{\lambda}x \right).
\end{equation}
Having computed $\alpha(x),\beta(x)$ the particular and quadrature solutions to Riccati equation are reconstructed by following Eq.\eqref{R6} as follows:
\begin{equation}\label{a7}
\begin{split}
        &y_p(x)=-\sqrt{\lambda}\exp\left(\frac{x^2}{2} \right),\\
        &y_q(x)=-\frac{\exp \left(\frac{x^2}{2}-2\sqrt{\lambda}x \right)}{C-\frac{1}{2\sqrt{\lambda}}\exp\left(-2\sqrt{\lambda}x\right)}.
\end{split}
\end{equation}
Then the general solution to this Riccati equation is given by:
\begin{equation}\label{a8}
    y(x)=-\sqrt{\lambda}\exp\left(\frac{x^2}{2} \right)-\frac{\exp \left(\frac{x^2}{2}-2\sqrt{\lambda}x \right)}{C-\frac{1}{2\sqrt{\lambda}}\exp\left(-2\sqrt{\lambda}x\right)}.
\end{equation}
As an additional example consider the following Riccati equation:
\begin{equation}\label{a9}
    \frac{dy}{dx}=-\frac{2}{x^2}\exp\left(\sin(x)\right)+\cos(x)y+\exp\left(-\sin(x) \right)y^2.
\end{equation}
From here, the following functions are indentified:
\begin{equation}\label{a10}
    q_0(x)=-\frac{2}{x^2}\exp\left(\sin(x)\right),\quad q_1(x)=\cos(x),\quad q_2(x)=\exp\left(-\sin(x) \right).
\end{equation}
From Eq.\eqref{a10} and by following Eq.\eqref{eq 2.1.14} $\beta(x)$ is computed:
\begin{equation}\label{a11}
    \beta(x)=0.
\end{equation}
In the same way, following Eq.\eqref{eq 2.1.18} an equation for $\alpha(x)$ is computed:
\begin{equation}\label{a12}
    \frac{d^2\alpha}{dx^2}=\frac{2}{x^2}\alpha.
\end{equation}
The solution for Eq.\eqref{a12} is given by Eq.\eqref{extra} as follows:
\begin{equation}\label{a13}
    \alpha(x)=x^2.
\end{equation}
Having computed $\alpha(x),\beta(x)$ the particular and quadrature solutions to Riccati equation are reconstructed by following Eq.\eqref{R6} as follows:
\begin{equation}\label{a14}
    \begin{split}
        &y_p(x)=-\frac{2}{x}\exp\left(\sin(x)\right),\\
        &y_q(x)=-\frac{\exp\left(\sin(x) \right)}{Cx^4-\frac{x}{3}}.
    \end{split}
\end{equation}
Then the general solution to this Riccati equation is given by:
\begin{equation}
    y(x)=-\frac{2}{x}\exp\left(\sin(x)\right)-\frac{\exp\left(\sin(x) \right)}{Cx^4-\frac{x}{3}}.
\end{equation}
\section{Conclusions}\label{6}
By employing the non-linear transformation detailed in Eq.\eqref{Riccati 2}, the Riccati equation was transformed into a second-order linear differential equation. This linear equation was subsequently solved using the generalized recursive integrating factors method as referenced in \cite{riveraoliva2025solvinglineardifferentialequations}. Upon resolution of the second-order differential equation, the application of the inverse transformation allowed for the reconstruction of the general solution to the Riccati equation, as indicated in Eq.\eqref{R4}. This process illustrates the efficacy of recursive methods in addressing differential equations, including those that are non-linear. Although Eq.\eqref{R4} provides a general solution formula for the Riccati equation, the convergence of this expression was not discussed. Consequently, despite its representation as a general formula, it may not converge to a solution in all cases. Further investigation into this aspect is warranted in future studies.


%
%

%

\begin{acknowledgments}
Everardo Rivera-Oliva wishes to express his gratitude for the support provided by the Ph.D. scholarship No. 743129 from SECIHTI (CONAHCYT)-Mexico.
\end{acknowledgments}
\bibliography{biblio}

\begin{thebibliography}{12}%
\makeatletter
\providecommand \@ifxundefined [1]{%
 \@ifx{#1\undefined}
}%
\providecommand \@ifnum [1]{%
 \ifnum #1\expandafter \@firstoftwo
 \else \expandafter \@secondoftwo
 \fi
}%
\providecommand \@ifx [1]{%
 \ifx #1\expandafter \@firstoftwo
 \else \expandafter \@secondoftwo
 \fi
}%
\providecommand \natexlab [1]{#1}%
\providecommand \enquote  [1]{``#1''}%
\providecommand \bibnamefont  [1]{#1}%
\providecommand \bibfnamefont [1]{#1}%
\providecommand \citenamefont [1]{#1}%
\providecommand \href@noop [0]{\@secondoftwo}%
\providecommand \href [0]{\begingroup \@sanitize@url \@href}%
\providecommand \@href[1]{\@@startlink{#1}\@@href}%
\providecommand \@@href[1]{\endgroup#1\@@endlink}%
\providecommand \@sanitize@url [0]{\catcode `\\12\catcode `\$12\catcode `\&12\catcode `\#12\catcode `\^12\catcode `\_12\catcode `\%12\relax}%
\providecommand \@@startlink[1]{}%
\providecommand \@@endlink[0]{}%
\providecommand \url  [0]{\begingroup\@sanitize@url \@url }%
\providecommand \@url [1]{\endgroup\@href {#1}{\urlprefix }}%
\providecommand \urlprefix  [0]{URL }%
\providecommand \Eprint [0]{\href }%
\providecommand \doibase [0]{http://dx.doi.org/}%
\providecommand \selectlanguage [0]{\@gobble}%
\providecommand \bibinfo  [0]{\@secondoftwo}%
\providecommand \bibfield  [0]{\@secondoftwo}%
\providecommand \translation [1]{[#1]}%
\providecommand \BibitemOpen [0]{}%
\providecommand \bibitemStop [0]{}%
\providecommand \bibitemNoStop [0]{.\EOS\space}%
\providecommand \EOS [0]{\spacefactor3000\relax}%
\providecommand \BibitemShut  [1]{\csname bibitem#1\endcsname}%
\let\auto@bib@innerbib\@empty
\bibitem [{\citenamefont {Riccati}(1724)}]{riccati1724animadversiones}%
  \BibitemOpen
  \bibfield  {author} {\bibinfo {author} {\bibfnamefont {J.}~\bibnamefont {Riccati}},\ }\href@noop {} {\bibfield  {journal} {\bibinfo  {journal} {Actorum Eruditorum Supplementa}\ }\textbf {\bibinfo {volume} {8}},\ \bibinfo {pages} {66} (\bibinfo {year} {1724})}\BibitemShut {NoStop}%
\bibitem [{\citenamefont {Nowakowski}\ and\ \citenamefont {Rosu}(2002)}]{PhysRevE.65.047602}%
  \BibitemOpen
  \bibfield  {author} {\bibinfo {author} {\bibfnamefont {M.}~\bibnamefont {Nowakowski}}\ and\ \bibinfo {author} {\bibfnamefont {H.~C.}\ \bibnamefont {Rosu}},\ }\href {\doibase 10.1103/PhysRevE.65.047602} {\bibfield  {journal} {\bibinfo  {journal} {Phys. Rev. E}\ }\textbf {\bibinfo {volume} {65}},\ \bibinfo {pages} {047602} (\bibinfo {year} {2002})}\BibitemShut {NoStop}%
\bibitem [{\citenamefont {Fraga}\ \emph {et~al.}(1999)\citenamefont {Fraga}, \citenamefont {de~la Vega},\ and\ \citenamefont {Fraga}}]{fraga1999schrödinger}%
  \BibitemOpen
  \bibfield  {author} {\bibinfo {author} {\bibfnamefont {S.}~\bibnamefont {Fraga}}, \bibinfo {author} {\bibfnamefont {J.}~\bibnamefont {de~la Vega}}, \ and\ \bibinfo {author} {\bibfnamefont {E.}~\bibnamefont {Fraga}},\ }\href {https://books.google.com.mx/books?id=hA3wAAAAMAAJ} {\emph {\bibinfo {title} {The Schr{\"o}dinger and Riccati Equations}}},\ Lecture Notes in Chemistry\ (\bibinfo  {publisher} {Springer Berlin Heidelberg},\ \bibinfo {year} {1999})\BibitemShut {NoStop}%
\bibitem [{\citenamefont {Schuch}(2014)}]{Schuch_2014}%
  \BibitemOpen
  \bibfield  {author} {\bibinfo {author} {\bibfnamefont {D.}~\bibnamefont {Schuch}},\ }\href {\doibase 10.1088/1742-6596/538/1/012019} {\bibfield  {journal} {\bibinfo  {journal} {Journal of Physics: Conference Series}\ }\textbf {\bibinfo {volume} {538}},\ \bibinfo {pages} {012019} (\bibinfo {year} {2014})}\BibitemShut {NoStop}%
\bibitem [{\citenamefont {Faraoni}(1999)}]{valerio}%
  \BibitemOpen
  \bibfield  {author} {\bibinfo {author} {\bibfnamefont {V.}~\bibnamefont {Faraoni}},\ }\href {\doibase 10.1119/1.19361} {\bibfield  {journal} {\bibinfo  {journal} {American Journal of Physics}\ }\textbf {\bibinfo {volume} {67}},\ \bibinfo {pages} {732} (\bibinfo {year} {1999})},\ \Eprint {http://arxiv.org/abs/https://pubs.aip.org/aapt/ajp/article-pdf/67/8/732/7528258/732\_1\_online.pdf} {https://pubs.aip.org/aapt/ajp/article-pdf/67/8/732/7528258/732\_1\_online.pdf} \BibitemShut {NoStop}%
\bibitem [{\citenamefont {Arfken}\ \emph {et~al.}(2013)\citenamefont {Arfken}, \citenamefont {Arfken}, \citenamefont {Weber},\ and\ \citenamefont {Harris}}]{arfken2013mathematical}%
  \BibitemOpen
  \bibfield  {author} {\bibinfo {author} {\bibfnamefont {G.}~\bibnamefont {Arfken}}, \bibinfo {author} {\bibfnamefont {G.}~\bibnamefont {Arfken}}, \bibinfo {author} {\bibfnamefont {H.}~\bibnamefont {Weber}}, \ and\ \bibinfo {author} {\bibfnamefont {F.}~\bibnamefont {Harris}},\ }\href {https://books.google.com.mx/books?id=qLFo_Z-PoGIC} {\emph {\bibinfo {title} {Mathematical Methods for Physicists: A Comprehensive Guide}}}\ (\bibinfo  {publisher} {Elsevier Science},\ \bibinfo {year} {2013})\BibitemShut {NoStop}%
\bibitem [{\citenamefont {Zill}(2016)}]{zill2016differential}%
  \BibitemOpen
  \bibfield  {author} {\bibinfo {author} {\bibfnamefont {D.}~\bibnamefont {Zill}},\ }\href {https://books.google.es/books?id=0NO5DQAAQBAJ} {\emph {\bibinfo {title} {Differential Equations with Boundary-Value Problems}}}\ (\bibinfo  {publisher} {Cengage Learning},\ \bibinfo {year} {2016})\BibitemShut {NoStop}%
\bibitem [{\citenamefont {Butkov}(1968)}]{butkov1968mathematical}%
  \BibitemOpen
  \bibfield  {author} {\bibinfo {author} {\bibfnamefont {E.}~\bibnamefont {Butkov}},\ }\href {https://books.google.com.mx/books?id=aGwNlA_i3pkC} {\emph {\bibinfo {title} {Mathematical Physics}}},\ A-W series in advanced physics\ (\bibinfo  {publisher} {Addison-Wesley Publishing Company},\ \bibinfo {year} {1968})\BibitemShut {NoStop}%
\bibitem [{\citenamefont {Rivera-Oliva}(2025)}]{riveraoliva2025solvinglineardifferentialequations}%
  \BibitemOpen
  \bibfield  {author} {\bibinfo {author} {\bibfnamefont {E.}~\bibnamefont {Rivera-Oliva}},\ }\href {https://arxiv.org/abs/2502.20219} {\enquote {\bibinfo {title} {Solving linear differential equations by recursion and integrating factors},}\ } (\bibinfo {year} {2025}),\ \Eprint {http://arxiv.org/abs/2502.20219} {arXiv:2502.20219 [math-ph]} \BibitemShut {NoStop}%
\bibitem [{\citenamefont {Hassani}(2013)}]{hassani2013mathematical}%
  \BibitemOpen
  \bibfield  {author} {\bibinfo {author} {\bibfnamefont {S.}~\bibnamefont {Hassani}},\ }\href {https://books.google.com.mx/books?id=uRa4BAAAQBAJ} {\emph {\bibinfo {title} {Mathematical Physics: A Modern Introduction to Its Foundations}}}\ (\bibinfo  {publisher} {Springer International Publishing},\ \bibinfo {year} {2013})\BibitemShut {NoStop}%
\bibitem [{\citenamefont {Borchardt}(1873)}]{+1873+214+235}%
  \BibitemOpen
  \bibinfo {editor} {\bibfnamefont {C.~W.}\ \bibnamefont {Borchardt}},\ ed.,\ \enquote {\bibinfo {title} {Ueber die integration der linearen differentialgleichungen durch reihen. von herrn g. frobenius},}\ in\ \href {\doibase doi:10.1515/9783112391525-016} {\emph {\bibinfo {booktitle} {Band 76}}}\ (\bibinfo  {publisher} {De Gruyter},\ \bibinfo {address} {Berlin, Boston},\ \bibinfo {year} {1873})\ pp.\ \bibinfo {pages} {214--235}\BibitemShut {NoStop}%
\bibitem [{\citenamefont {Grossman}\ and\ \citenamefont {Katz}(1972)}]{grossman1972non}%
  \BibitemOpen
  \bibfield  {author} {\bibinfo {author} {\bibfnamefont {M.}~\bibnamefont {Grossman}}\ and\ \bibinfo {author} {\bibfnamefont {R.}~\bibnamefont {Katz}},\ }\href {https://books.google.com.mx/books?id=RLuJmE5y8pYC} {\emph {\bibinfo {title} {Non-Newtonian Calculus: A Self-contained, Elementary Exposition of the Authors' Investigations ...}}}\ (\bibinfo  {publisher} {Lee Press},\ \bibinfo {year} {1972})\BibitemShut {NoStop}%
\end{thebibliography}%

\end{document}